%% file: subset-fvs.tex
\documentclass{article}

\usepackage{fullpage}

\usepackage[
normalsections, 
normalmargins, 
normallists, 
normalfloats, 
normalindent, 
normaltitle, 
normalleading, 
normallooseness, 
]{savetrees}

\usepackage{amsmath,amstext,amssymb,amsthm,comment}
\usepackage{tikz}
\usepackage{pgflibrarysnakes}
\usetikzlibrary{snakes}
\usepackage{times}
\usepackage{calc}
\usepackage{amsopn}
\usepackage[noend]{algorithmic}
\usepackage[boxed]{algorithm}
\usepackage{latexsym}

\newtheorem{theorem}{Theorem}[section]
\newtheorem{lemma}[theorem]{Lemma}
\newtheorem{definition}[theorem]{Definition}

\newtheorem{reduction}{Reduction}
\newtheorem{step}{Step}

\newcommand{\N}{\ensuremath{\mathbb{N}}}

\floatname{algorithm}{Algorithm}

\newcommand{\defproblemu}[4]{
  \vspace{1mm}
\noindent\fbox{
  \begin{minipage}{0.95\textwidth}
  \begin{tabular*}{\textwidth}{@{\extracolsep{\fill}}lr} #1 & {\bf{Parameter:}} #3 \\ \end{tabular*}
  {\bf{Input:}} #2  \\
  {\bf{Question:}} #4
  \end{minipage}
  }
  \vspace{1mm}
}

\newcommand{\dirfvsname}{{\sc{Directed Feedback Vertex Set}}}
\newcommand{\fvsname}{{\sc{Feedback Vertex Set}}}
\newcommand{\sfvsname}{{\sc{Subset Feedback Vertex Set}}}
\newcommand{\sfesname}{{\sc{Edge Subset Feedback Vertex Set}}}

\newcommand{\sfvsshort}{{\sc{Subset-FVS}}}
\newcommand{\esfvsshort}{{\sc{Edge-Subset-FVS}}}
\newcommand{\sfesshort}{{\sc{Edge-Subset-FVS}}}
\newcommand{\fvsshort}{{\sc{FVS}}}

\newcommand{\multiwayname}{{\sc{Multiway Cut}}}

\newcommand{\nodemultiwayname}{{\sc{Node Multiway Cut}}}
\newcommand{\multicutname}{{\sc{Multicut}}}
\newcommand{\nodemulticutname}{{\sc{Node Multicut}}}

\newcommand{\TT}{ {\mathcal{T} }}
  
\title{Subset feedback vertex set is fixed-parameter tractable\thanks{A preliminary version of this paper was presented at the 38th International Colloquium on Automata, Languages and Programming,
Z\"{u}rich, Switzerland, 2011.}}

  \author{
      Marek Cygan\thanks{Institute of Informatics, University of Warsaw, Poland, e-mail: \texttt{cygan@mimuw.edu.pl}. Partially supported by Foundation for Polish Science and Polish Ministry of Science grant no. N206 491238}
    \and
      Marcin Pilipczuk\thanks{Institute of Informatics, University of Warsaw, Poland, e-mail: \texttt{malcin@mimuw.edu.pl}. Partially supported by Foundation for Polish Science and Polish Ministry of Science graph no. N206 491038}
    \and
      Micha\l{} Pilipczuk\thanks{Faculty of Mathematics, Informatics and Mechanics, University of Warsaw, Poland, e-mail: \texttt{mp248287@students.mimuw.edu.pl}}
    \and
      Jakub Onufry Wojtaszczyk\thanks{Institute of Mathematics, University of Warsaw, Poland and Google Inc., Cracow, Poland, e-mail: \texttt{onufry@google.com}}
  }

\begin{document}

\maketitle

\begin{abstract}

	The classical \fvsname{} problem asks, for a given undirected graph $G$
and an integer $k$, to find a set of at most $k$ vertices that hits all the
cycles in the graph $G$.
	\fvsname{} has attracted a large amount of research in the parameterized
setting, and subsequent kernelization and fixed-parameter algorithms have
been a rich source of ideas in the field.

	In this paper we consider a more general and difficult version of the
problem, named \sfvsname{} (\sfvsshort{} in short) where an instance comes 
additionally with a set $S \subseteq V$ of vertices, and we ask for a set 
of at most $k$ vertices that hits all simple cycles passing through $S$.
Because of its applications in circuit testing and genetic linkage analysis
\sfvsshort{} was studied from the approximation algorithms perspective
by Even et al.~[SICOMP'00, SIDMA'00].

	The question whether the \sfvsshort{} problem is fixed-parameter tractable was posed
independently by Kawara-bayashi and Saurabh in 2009.
	We answer this question affirmatively.
	We begin by showing that this problem is fixed-parameter tractable when
parametrized by $|S|$.
	Next we present an algorithm which reduces the given instance
  to $2^k n^{O(1)}$ instances with the size of $S$ bounded by $O(k^3)$,
  using kernelization techniques such as
the $2$-Expansion Lemma, Menger's theorem and Gallai's theorem.
	These two facts allow us to give a $2^{O(k\log k)} n^{O(1)}$ time algorithm
solving the \sfvsname{} problem, proving that it is indeed fixed-parameter
tractable.
\end{abstract}

\input{r1-introduction}

\input{r3-small-set}

\input{r4-general-case}

\input{r-reduction}

\input{r-conclusions}

\bibliographystyle{plain}
\bibliography{subset-fvs}

\end{document}

%% file: r1-introduction.tex
\newcommand{\cN}[1]{\ensuremath{N[#1]}}

\section{Introduction} \label{sec:intro}

\fvsname{} (\fvsshort{}) is one of the long--studied problems in the 
algorithms area.
It can be stated as follows: given an undirected graph $G$ on $n$ vertices 
and a parameter $k$ decide if one can remove at most $k$ vertices from $G$ 
so that the remaining graph does not contain a cycle, i.e., is a forest.
The problem of finding feedback sets in undirected graphs
arises in a variety of applications in genetics, circuit testing,
artificial intelligence, deadlock resolution,
and analysis of manufacturing processes \cite{even:sfvs}.

Because of its importance the feedback vertex set problem was studied 
from the approximation algorithms perspective in
different variants and generalisations including \dirfvsname{}
and \sfvsname{} (see~\cite{dirfvs:apx} and~\cite{sfvs:8-apx} for further 
references).
In this paper we will study the \sfvsname{} problem from the parametrized
complexity perspective.

In the parameterized complexity setting, an instance comes with an integer 
parameter $k$ --- formally, a parameterized problem $Q$ is a subset of 
$\Sigma^* \times \N$ for some finite alphabet $\Sigma$.
We say that a problem is {\em{fixed-parameter tractable}} ({\em{FPT}}) 
if there exists an algorithm solving any instance $(x,k)$ in time $f(k) 
{\rm poly}(|x|)$ for some (usually exponential) computable function $f$.
Intuitively, the parameter $k$ measures the hardness of the instance.
Fixed-parameter tractability has received much notice
as a method of effectively solving NP-hard problems for instances 
with a small parameter value.

The long line of research concerning \fvsshort{} 
in the parameterized complexity setting contains 
\cite{fvs:4krand,fvs1,fvs:3.83k,fvs:5k,fvs7,fvs2,fvs3,guo:fvs,fvs6,fvs5}.
Currently the fastest known algorithm works in $3^k n^{O(1)}$ time 
\cite{fvs:3k}.
Thomass\'{e}~\cite{fvs:quadratic-kernel} has shown a quadratic kernel for 
this problem improving previous results~\cite{fvs:kernel2,fvs:kernel1}.
The directed version has been proved to be FPT in 2008 by Chen et 
al.~\cite{directed-fvs}, closing a long-standing open problem in the 
parameterized complexity community. 
The natural question concerning the parameterized complexity of the 
\sfvsname{} problem was posed independently by Kawarabayashi at the 4th 
workshop on Graph Classes, Optimization, and Width Parameters (GROW 2009) 
and by Saurabh at the Dagstuhl seminar 09511~\cite{dagstuhl:2009}.

\paragraph{Notation}
Let us now introduce some notation.
Let $G = (V,E)$ be a simple undirected graph with $n$ vertices.
A {\em cycle} in $G$ is a sequence of vertices $v_1v_2\ldots v_m \in V$
such that $v_iv_{i+1} \in E$ and $v_mv_1 \in E$.
We say a cycle is {\em simple} if $m>2$ and the vertices $v_i$ are pairwise 
different.
We will also consider multigraphs (i.e., graphs with multiple edges and 
loops), in which a simple cycle can have two vertices if there is a 
multiple edge between them, or a single vertex if there is a loop attached to it.
We call an edge $vw \in E$ a {\em bridge} if in $(V, E\setminus \{vw\})$
the vertices $v$ and $w$ are in different connected components.
Note that no simple cycle can contain a bridge as one of its edges.
Given subsets $X,Y \subseteq V$, by $E(X,Y)$ we denote the set of edges 
with one endpoint in $X$ and the other in $Y$.
By $G[X]$ we denote the subgraph induced by $X$ with the edge set $E(X,X)$.
By $N(X)$ we denote the neighbourhood of $X$, i.e. 
$\{u\in V \setminus X: \exists_{v\in X} uv\in E\}$.
For a subset of edges $E'\subseteq E$ by $V(E')$ we denote the set of all 
endpoints of edges from the set $E'$.

\paragraph{Problem definitions}
In this paper we study the \sfvsname{} problem (\sfvsshort{}), 
where an instance comes with a subset of vertices $S$, and we ask for a set 
of at most $k$ vertices that hits all simple cycles passing through $S$.
It is easy to see that \sfvsshort{} is a generalisation of \fvsshort{}
by putting $S=V$.
The weighted version of \sfvsshort{} was introduced by Even et al. 
\cite{even:sfvs} as a generalization of two problems:
\fvsname{} and \nodemultiwayname{}.
Even et al. motivate \sfvsshort{} problem by
explaining its applicability to genetic linkage.

\defproblemu{\sfvsname{} (\sfvsshort)}{An undirected graph $G = (V,E)$, 
a set $S \subseteq V$ and a positive integer $k$}{$k$}
{Does there exist a set $T \subseteq V$ such that $|T| \leq k$ and 
no simple cycle in $G[V\setminus T]$ contains a vertex of $S$?}

We also define a variant of \sfvsshort{}, where the set $S$ is a subset of 
edges of $G$.

\defproblemu{\sfesname{} (\sfesshort)}{An undirected graph $G = (V,E)$, 
a set $S \subseteq E$ and a positive integer $k$}{$k$}
{Does there exist a set $T \subseteq V$ with $|T| \leq k$, 
such that no simple cycle in $G[V\setminus T]$ contains an edge from $S$?}

The two problems stated above are equivalent. 
To see this, note that if $(G,S,k)$ is an instance of \sfvsshort{}, 
we create an instance $(G, S', k)$ of \sfesshort{}
by selecting as $S'$ all the edges incident to any vertex of $S$.
Then any simple cycle passing through a vertex of $S$ has to pass through
an edge of $S'$, and conversely, any cycle passing through an edge of $S'$
contains a vertex from $S$.
In the other direction, if $(G,S',k)$ is an instance of \sfesshort{}, 
obtain $G'$ by replacing
each edge $uv \in S'$ by a path $u-x_{uv}-v$ of length $2$, and solve
the \sfvsshort{} instance $(G',S,k)$ where $S = \{x_e: e \in S'\}$.
Clearly both reductions work in polynomial time and do not change the 
parameter.
Thus, in the rest of this paper we focus on solving \sfesname{}.
A simple cycle containing an edge from $S$ is called an $S$--{\em{cycle}}.

Let us recall here the definitions of two other problems related to \sfvsshort{}.

\defproblemu{\nodemultiwayname{}}{An undirected graph $G=(V,E)$,
  a set of vertices $\mathcal{T} \subseteq V$, called {\em{terminals}},
  and a positive integer $k$}{$k$}{Does there exist a set $T \subseteq V$
of at most $k$ non-terminals, such that no two terminals are in the same connected component of $G[V \setminus T]$?}

\defproblemu{\nodemulticutname{}}{An undirected graph $G=(V,E)$,
  a set of pairs of vertices $\mathcal{T} \subseteq V \times V$, called {\em{terminal pairs}},
  and a positive integer $k$}{$k$}{Does there exist a set $T \subseteq V$ of at most $k$ non-terminals, such that no terminal pair is contained in one connected component of $G[V \setminus T]$?}

\paragraph{Our contributions} The main result of the paper is the following.

\begin{theorem}\label{thm:main}
  There exists a $2^{O(k \log k)} n^{O(1)}$-time and polynomial space algorithm for \sfesshort{}
(which implies an algorithm of the same time complexity for \sfvsshort{}).
\end{theorem}

This result resolves an open problem posted in 2009
independently by Kawarabayashi and by Saurabh.
To achieve this result we use several tools such as iterative compression,
the $2$-Expansion Lemma, Menger's theorem, Gallai's theorem and the 
algorithm for the \multiwayname{} problem.
Some of our ideas were inspired by previous FPT results: the algorithm 
for \multicutname{}
parameterized by $(|\TT|,k)$ by Guillemot~\cite{guill:8tk}, 
the $37.7^k n^{O(1)}$--time 
algorithm for \fvsshort{} by Guo et al.~\cite{guo:fvs} and
the quadratic kernel for \fvsshort{} 
by Thomass\'{e}~\cite{fvs:quadratic-kernel}.

We do not analyze the value of the exponent in the term $n^{O(1)}$, as
in our algorithm it is far from being linear.
The most important reasons for this dependency on $n$ is that the usage of Gallai's theorem
requires finding of a maximum matching in an auxiliary graph, and the use of
iterative compression gives additional multiplicative factor of $n$.

\paragraph{Related work}
As observed by Even et al.~\cite{even:sfvs} the weighted version 
of \sfvsshort{} is a generalisation of \nodemultiwayname{}. 
It is straightforward to adjust their reduction to the unweighted parameterized case;
for sake of completeness, we include the reduction in Section \ref{sec:red}.

Recently a lot of effort was put into developing kernelization and FPT
algorithms for terminal separation problems, including the quadratic kernel
\cite{fvs:quadratic-kernel} and the fast FPT algorithm \cite{fvs:3k}
for \fvsshort{} and the results resolving the parametrized complexity
status of \multicutname{} independently obtained by Bousquet et 
al.~\cite{multicut2} and by Marx and Razgon~\cite{multicut1}.
To the best of our knowledge, though, none of those results implies an FPT 
algorithm for \sfvsshort{}.

\sfvsshort{} was studied from the approximation perspective and the best 
known approximation algorithm by Even at al.~\cite{sfvs:8-apx}
gives approximation ratio equal to $8$.

We were recently informed that an FPT algorithm for \sfvsshort{} was 
independently discovered by Kawarabayashi and Kobayashi \cite{kebab:sfvs}.
Their algorithm uses significantly different techniques (minor theory) and 
its dependency on $k$ in the running time is worse than $2^{O(k \log k)}$.

\paragraph{Outline of the paper}
In Section~\ref{sec:small} we present an FPT algorithm for \esfvsshort{} 
when parameterized
by $|S|$. For the sake of presentation, we first give an easy-to-describe 
$f(|S|)n^{O(1)}$ algorithm at the cost of a fast growing function $f$ (Section \ref{sec:small1}).
Then we enhance this algorithm using techniques of Guillemot \cite{guill:8tk}
so that the function $f$ is replaced by $2^{O(k\log |S|)}$ (Section \ref{sec:small2}).
Later, in Section~\ref{sec:final} we develop an algorithm that
produces $2^k n^{O(1)}$ subinstances with the size of $S$ bounded by $O(k^3)$.
In Section \ref{sec:red} we include a reduction from \nodemultiwayname{}
to \sfvsshort{} in the parameterized setting.
Finally, Section \ref{sec:conc} contains conclusions and open problems.

%% file: r3-small-set.tex
\newcommand{\PP}{ {\mathcal{P} }}
\section{\esfvsshort{} parameterized by $|S|$}\label{sec:small}

In this section we concentrate on solving the \sfesshort{} problem parameterized by $|S|$, which
means that our complexity function can be exponentially dependent on the number
of edges in the set $S$.
This is the first step towards obtaining an FPT algorithm when parameterized by $k$.
Observe that we may assume $k < |S|$ since otherwise we may delete one vertex from
each edge from the set $S$ thus removing all edges from the set $S$ from our graph.

Let us first introduce some notation.
For $G = (V,E)$ denote $G_S=(V, E\setminus S)$.
By a {\em partition} of a set $Z$ we mean such a family $\PP = \{P_1, \ldots, P_m\}$, that $P_i$s are pairwise disjoint and their union is $Z$.
We say a partition $\PP'$ is a {\em subpartition} of $\PP$ if every element of $\PP'$ is contained in some element of $\PP$, in this case we call $\PP$ a {\em superpartition} of $\PP'$.

\subsection{Simpler and slower algorithm}\label{sec:small1}

We begin by showing an FPT algorithm which is easy to understand and later we present methods
to improve the time complexity.
We use the fact that \nodemulticutname{} is FPT when
parameterized by $(k,|\TT|)$ which was shown by Marx~\cite{marx:multiway-journal}.

\begin{theorem}
  There exists an algorithm solving the \sfesname{} problem in $f(|S|)n^{O(1)}$ time, for
  some computable function $f$.
\end{theorem}

\begin{proof}
Let $T$ be some solution of \sfesshort{}.
Our new parametrization, by $|S|$, allows us to guess, by checking all possibilities,
the subset $T_S = T \cap V(S)$ that is removed by the solution $T$.
Moreover, our algorithm guesses how the set $V(S) \setminus T_S$
is partitioned into connected components in the graph $G_S[V \setminus T]$.
Clearly both the number of subsets and of possible partitions is a function of $|S|$.
For a partition $\PP = \{P_1,\ldots,P_m\}$ of $V(S)\setminus T_S$ we form a multigraph $G_\PP$ on the set 
$\{P_1,\ldots,P_m\}$ by adding an edge $P_iP_j$ for every edge $uv \in S$,
where $u\in P_i, v\in P_j$.
Now we check whether there exists an edge in $G_{\PP}$ which is not a bridge.
If that is the case we know that the partition $\PP$ does not correspond to any solution
of \sfesshort, as any simple cycle in $G_{\PP}$ can be converted into a simple cycle in $G$ ---
hence we skip this partition.
Otherwise we create a set of pairs $\TT$, 
containing all pairs of vertices from the set $V(S) \setminus T_S$
that belong to different sets in the partition $\PP$.
Formally $\TT = \{(v_i,v_j) : v_i \in P_{i'}, v_j \in P_{j'}, i' \not = j'\}$.
Because of the properties of the multigraph $G_{\PP}$ 
it is sufficient to ensure that no pair from the set $\TT$
is contained in one connected component, hence
the last step is calling an algorithm for the \nodemulticutname{} problem with parameter $k - |T_S|$.
If the call returns a positive answer and a solution $X$,
the set $T_S\cup X$ is a solution to \sfesshort{}:
the connected components of $G_S[V \setminus (T_S \cup X)]$ induce a partition of $V(S)\setminus (T_S \cup X)$ that is a subpartition of $\PP$
and thus all remaining edges of $S$ are bridges in $G[V \setminus (T_S \cup X)]$.
Note that we do not require here that $X \cap V(S) = \emptyset$ nor that
the induced partition of $V(S) \setminus (T_S \cup X)$ is exactly the partition $\PP$ (being
a subpartition is sufficient).
On the other hand, if the answer to \sfesshort{} is positive, the \nodemulticutname{} call returns a solution for at least one choice of $T_S$ and $\PP$, the one implied by the \sfesshort{} solution.
Observe that $|\TT|=O(|S|^2)$ so we obtain an FPT algorithm for the \sfesname{} problem
parameterized by $|S|$.\end{proof}

{\small
      \begin{center}
      \fbox{
      \begin{minipage}{0.95\textwidth}
      \begin{algorithmic}[1]
      \REQUIRE $\mathrm{EdgeSubsetFeedbackVertexSet}(G,S,k)$
      \COMMENT{parameterized by $(|S|,k)$}
      \FORALL{subsets $T_S\subseteq V(S), |T_S| \le k$}
        \FORALL{partitions $\PP=\{P_1,\ldots,P_m\}$ of $V(S) \setminus T_S$}
          \STATE form a multigraph $G_{\PP}$ on the set $\{P_1,\ldots,P_m\}$ by adding an edge $P_iP_j$
            for every edge $uv \in S$, $u\in P_i$, $v\in P_j$.
          \IF {all edges in $G_{\PP}$ are bridges}
            \STATE let $\TT=\{(v_i,v_j):v_i \in P_{i'}, v_j\in P_{j'}, i' \not = j'\}$ 
            \IF{MultiCut$(G_S[V\setminus T_S],\TT,k-|T_S|)$ returns $(YES,X)$ }
              \STATE {\bf return $T_S \cup X$} 
            \ENDIF
          \ENDIF
        \ENDFOR
      \ENDFOR
          
      \STATE {\bf return} $NO$
      \end{algorithmic}
      \end{minipage}
      }
      \end{center}
}

\subsection{Improving the time complexity}\label{sec:small2}
Our whole approach in this subsection is closely based to the arguments of Guillemot~\cite{guill:8tk} for \multicutname.
We first recall that \nodemultiwayname{} is fixed-parameter tractable when parameterized
by the solution size $k$, and currently the best running time is $2^k n^{O(1)}$
\cite{nmc:2k}.

Our main result in this section is the following:
\begin{theorem}\label{thm:small-alg}
There exists an algorithm solving the \sfesname{} problem in
$2^{O(k\log |S|)}n^{O(1)}$ time.
\end{theorem}

\begin{proof}
The algorithm works in three phases.
In the first two phases we aim to divide the set of all endpoints of edges from $S$ into a family of subsets.
We prove that for each set $T$ of at most $k$ vertices from $V$ there exists a generated partition of $V(S)$
that is the same as the partition of $V(S)$ induced by the connected components of $G_S[V \setminus T]$.
In the third phase we check whether we can, in fact, achieve a generated partition of the endpoints of edges from $S$ into connected components by removing at most $k$ vertices, and whether such a partition implies that we removed all cycles passing through $S$ from $G$.

Initialize $R = \emptyset$. The first phase works as follows:
\begin{enumerate}
\item Select a spanning forest $F$ of $G_S[V \setminus R]$, let $U = V(S) \setminus R$ be the set of endpoints of edges from $S$ outside $R$;\label{punkt:1}
\item If $k = |R|$ proceed directly to phase two;
\item As long as there are isolated vertices not from $U$ or leaves not from $U$ in $F$, remove them from $F$;
\item As long as there are vertices of degree 2 not from $U$ in $F$, remove them from $F$, and connect the two neighbours of the removed vertex with an edge in $F$;\label{punkt:4}
\item Branch out --- one branch passes the resultant forest $F$ and sets $U$ and $R$ to the second phase. In other $|F|$ branches we select a vertex from $F$, add it to $R$ and go back to Point \ref{punkt:1}.
\end{enumerate}

Note that after the first four steps of phase one $F$ has at most $2|V(S)|$ vertices (as all its vertices of degree at most $2$ are from $U \subseteq V(S)$).
Thus in the fifth step we have at most $2|V(S)| + 1$ branches.
As we can branch out into phase one at most $k$ times due to steps 2 and 5, this assures we have at most $(2|V(S)| + 1)^k$ entries into phase two from phase one.
The internal workings of the first phase are obviously polynomial--time.

The second phase is somewhat more complicated, and aims at arriving at a partition of the set $U = V(S) \setminus R$, based on the forest $F$ received from phase one.
Informally speaking, the set $R$ is to be included in the solution and in the second phase we choose a subgraph of $F$ that corresponds to connected components of $G_S[V \setminus T]$.

Let $\PP = \{P_1, \ldots, P_m\}$ be the partition of $U$ given by phase one --- that is if we denote the connected components of $F$ by $C_1, C_2, \ldots, C_m$, then $P_i = C_i \cap U$
for $1 \leq i \leq m$.
Note that the $P_i$s are non--empty, since if some connected component of $F$ contained no vertex from $U$, then it would be removed from $F$ completely during the third step of phase one.
Now we proceed as follows:
\begin{enumerate}
\item Select at most $k - |R|$ edges from $F$;
\item Let $\PP'$ denote the partition of $U$ given (as above) by $F$ with the selected edges removed;
\item For each partition $\PP''$ of $U$ being at the same time a subpartition of $\PP$ and a superpartition of $\PP'$ start phase three.
\end{enumerate}

We want to check how many times each application of phase two enters phase three.
As $F$ has at most $2|U|\leq 2|V(S)|$ edges, we have at most $(2|V(S)|)^k$ choices in the first step.
Now, consider a connected component $C_i$ of $F$, from which we removed $s_i$ edges.
There are exactly $s_i+1$ elements of $\PP'$ which are subsets of $C_i$.
The number of ways to combine these elements into a partition of $C_i \cap U$ is the $(s_i+1)$--st Bell number $B_{s_i+1} \leq (s_i+1)^{s_i+1}$.
The product of these is
\begin{align*}
  \prod_i B_{s_i + 1} &\leq \prod_i (s_i + 1)^{s_i + 1} =
  \exp \left(\sum_i (s_i + 1) \log (s_i + 1)\right) \\
  &\leq \exp \left(\left(1 + \sum_i s_i\right) \log \left(1 + \sum_i s_i\right)\right) = (1 + k)^{1 + k},
\end{align*} 
where the last inequality follows from the standard Jensen's inequality corollary $f(\sum_i x_i) \leq \sum_i f(x_i)$ for a convex function $f$ with $f(0) = 0$ and non--negative $x_i$ applied for $f(x) = (1+x)\log (1+x)$.
Thus for each execution of phase two, phase three is executed at most $(2|V(S)|)^k (k+1)^{k+1}$ times.

Phase three works as follows:
\begin{enumerate}
\item Take the partition $\PP''$ of $U = V(S)\setminus R$ given by phase two, form a multigraph on the set $\{P_1, P_2, \ldots, P_m\}$ by adding an edge $P_iP_j$ for every edge $uv \in S$, where $u \in P_i$, $v \in P_j$. If any of these edges is not a bridge, return a negative answer from this branch;
\item Otherwise create a graph $G'$ by adding to $G$ a vertex $w_i$ for every $P_i \in \PP''$, $i \geq 1$ and adding edges connecting $w_i$ to $u_j$ for each $u_j \in P_i$;
let $W = \{w_i\}_{i = 1}^m$;
\item Apply \nodemultiwayname{} to $(G_S'[(V \cup W) \setminus R], W, k - |R|)$.
If it returns a negative answer, we return a negative answer from this branch, otherwise we return $Q \cup R$ as a solution, where $Q$ is the set returned by \nodemultiwayname.
\end{enumerate}

The execution of this phase takes $2^k n^{O(1)}$ time \cite{nmc:2k}.
Thus, as the first phase branches out into at most $(2|V(S)| + 1)^k$ executions of phase two, phase two branches out into at most $(2|V(S)|)^k (k+1)^{k+1}$ instances of phase three, and phase three executes in $2^k n^{O(1)}$, the runtime of the whole algorithm is at most $2^{O(k \log |V(S)|)} n^{O(1)}$ (we may assume $k \leq |V(S)|$ --- if otherwise, removing $V(S)$ from $V$ gives a trivial positive solution).

Now we prove the correctness of this algorithm.
First assume the algorithm returns a solution.
It was then found by phase three.
Consider the graph $G'[(V \cup W) \setminus (Q \cup R)]$.
As $Q$ was returned by the \nodemultiwayname{} algorithm, we know $|Q| \leq k - |R|$ (and thus $|Q \cup R| \leq k$).
We prove there are no simple cycles through edges of $S$ in $G'[(V \cup W) \setminus (Q \cup R)]$, which implies the same for its subgraph $G[V \setminus (Q \cup R)]$.
As $Q$ was returned by \nodemultiwayname, each vertex $w_i$ is in a distinct connected component of $G_S'[(V\cup W) \setminus (Q \cup R)]$.
Consider a simple cycle in $G'[(V \cup W) \setminus (Q \cup R)]$ passing through some edge $e \in S$.
This means there is a path connecting the two endpoints of $e$ in $G'[(V\cup W) \setminus (Q \cup R)]$ not passing through $e$.
We may contract connected components containing the $w_i$s to single vertices, thus receiving a graph isomorphic to the graph considered in the first step of phase three.
The existence of the simple path, however, means $e$ was not a bridge in this graph, contrary to the assumption our algorithm returned $Q \cup R$ as a solution.
Thus any solution found by our algorithm is indeed a solution of the \sfesshort{} problem.

On the other hand, assume there exists some solution $T$ of the \sfesshort{} problem.
We show how our algorithm arrives at a positive answer in this case.
In the first phase, if any vertex $t \in T$ remains in the forest $F$ constructed in this phase in Point \ref{punkt:4}, we select the branch that adds $t$ to $R$.
If no vertex from $T$ remains in $F$ at some iteration of the first phase, we branch out to the second phase. Note that $R \subseteq T$.
In the second phase, we choose the edges into which the vertices from $T$ were contracted (if some were dropped due to being leaves or isolated vertices in $F$, we simply choose fewer edges).
Now consider the partition $\PP''$ of $U = V(S)\setminus R$ given by the relation of being in the same connected component of $G_S[V \setminus T]$.
It is a subpartition of $\PP$, as $\PP$ is simply the partition given by $G_S[V \setminus R]$, and $R \subseteq T$.
It is also a superpartition of $\PP'$, as $\PP'$ is given by removing the whole $T$ and all the edges which are not edges of the forest $F$.
Thus $\PP''$ is one of the partitions considered in the second phase, and thus enters the third phase.
Now for this partition the edges of $S$ are bridges in the sense of the first step of phase three, as the vertices of the graph considered there correspond exactly to connected components of $G_S[V\setminus T]$, and $G[V \setminus T]$ has no simple cycles through edges in $S$.
Moreover, $T \setminus R$ is a solution for the \nodemultiwayname{} problem by its definition.
Thus \nodemultiwayname{} returns some positive answer (not necessarily $T \setminus R$) in this branch, and thus the algorithm gives the correct answer.
\end{proof}

%% file: r4-general-case.tex
\newcommand{\Dd}{ {\mathcal{D} }}

\newcommand{\dsfes}{{\sc{Disjoint}} \sfesshort{}}
\section{\sfesshort{} parameterized by $k$}\label{sec:final}

In this section we show an FPT algorithm for \sfesname{}.

We begin by noting that, using standard arguments, one can show that \sfesshort{} is {\em self-reducible} --- i.e., if we have an algorithm that solves \sfesshort,
we can also find a witness: a set $T$ that intersects all cycles passing through $S$.
The procedure is standard: Assume the answer is positive.
For every vertex we check whether it can be a part of the solution by removing it
from the graph, decreasing $k$ by one and running our algorithm on the reduced instance.
For at least one vertex the answer has to be positive, we greedily take any such vertex into the solution
and proceed inductively.

We now follow the idea of {\em{iterative compression}} proposed by Reed et al. \cite{reed:ic}.
First, note that if $V' \subseteq V$ and $T$
is a feasible solution to an \sfesshort{} instance $(G, S, k)$, then $V' \cap T$ is a feasible
solution to the instance $(G[V'], S', k)$, where $S' = S \cap E(G[V'])$.
Thus, if the answer for $(G[V'], S', k)$ is negative,
so is the answer for $(G, S, k)$.
Let $V = \{v_1, v_2, \ldots, v_n\}$ be an arbitrary ordering of the set of vertices of $G$.
We consecutively construct solutions to \sfesshort{} for instances $\mathcal{I}_i = (G[V_i], S_i, k)$, where
$V_i = \{v_1, v_2, \ldots, v_i\}$ and $S_i = S \cap E(G[V_i])$.
When looking for a solution for graph $G[V_{i+1}]$, we use the fact that if $T_i$ is a
solution for $\mathcal{I}_i$, then $Z_{i+1} = T_i \cup \{v_{i+1}\}$ is a solution
for $(G[V_{i+1}], S_{i+1}, k+1)$ --- a solution for our problem with the parameter increased by one.

We start with a standard branching into $2^{|Z_{i+1}|}$ subcases, guessing which vertices from $Z_{i+1}$
are taken into a solution to the instance $\mathcal{I}_{i+1}$. Let us focus on a fixed branch,
where we decided to take $T_Z \subseteq Z_{i+1}$ into a solution
and denote $Z = Z_{i+1} \setminus T_Z$.
We delete $T_Z$ from the graph $G$, reduce $S$ to $S \cap E(G \setminus T_Z)$, and decrease $k$ by $|T_Z|$, arriving at the following subproblem.

\defproblemu{\dsfes{}}{A \sfesshort{} instance $(G,S,k)$ together with a set $Z \subseteq V(G)$ that is a solution to the \sfesshort{} instance $(G,S,|Z|)$}{$k$ and $|Z|$}{Does there exist a solution to $(G,S,k)$ that is disjoint with $Z$?}

However, we are not going to provide an algorithm that solves any \dsfes{} instance, but
only a {\em{maximal}} one. Informally speaking, we are only interested in those of
$2^{|Z_{i+1}|}$ branches, where the guessed set $T_Z$ is (inclusion-wise) maximal. Formally:

\begin{definition}\label{def:maximal}
We say that a \dsfes{} instance $(G,S,k,Z)$ is a {\em{maximal}} instance if every feasible solution to \sfesshort{} instance  $(G,S,k)$ is disjoint with $Z$.
\end{definition}

We provide a set of reductions that reduce the size of $S$ to polynomial in $k$.
However, we do not require that the reductions are sound with respect to any \dsfes{} instance,
but only to the maximal ones. Formally, we define the reductions as follows.

\begin{definition}\label{def:reduce}
We say that a \dsfes{} instance $(G',S',k',Z')$ is a {\em{properly reduced}} instance $(G,S,k,Z)$ if the following holds:
\begin{enumerate}
\item $|V(G')| \leq |V(G)|$ and $k' \leq k$;
\item if $(G,S,k)$ is a \sfesshort{} NO-instance, so is $(G',S',k')$;
\item if $(G,S,k,Z)$ is a maximal \dsfes{} YES-instance, so is $(G',S',k',Z')$.
\end{enumerate}
\end{definition}

We are now ready to state the main theorem of this section.

\begin{theorem}\label{thm:reduce}
There exists a polynomial-time algorithm $\mathcal{R}$
that, given a \dsfes{} instance $(G,S,k,Z)$, either:
\begin{enumerate}
\item returns a properly reduced instance $(G',S',k',Z')$ with $k' \leq k$, $|Z'| \leq |Z|$ and $|S'| = O(k'|Z'|^2)$;
\item or returns IGNORE, in this case $(G,S,k,Z)$ is not a {\em{maximal}} \dsfes{} YES-instance.
\end{enumerate}
\end{theorem}

We first show that Theorem \ref{thm:reduce} leads to the desired FPT algorithm for \sfesshort{},
i.e., we now prove Theorem \ref{thm:main}.

{\em{Proof of Theorem \ref{thm:main}}}. 
In each step of the iterative compression, in each of $2^{|Z_{i+1}|}$ branches, we run the algorithm $\mathcal{R}$.
If it gives the second answer, we ignore this branch.
In case of the first answer, we invoke the algorithm from Theorem \ref{thm:small-alg} on \sfesshort{} instance $(G',S',k')$, leading to running time $2^{O(k \log k)} n^{O(1)}$.
Note that if $(G',S',k')$ is a \sfesshort{} YES-instance, so is $(G,S,k)$ (by the second property of Definition \ref{def:reduce}),
and any solution (even not disjoint with $Z$) to $(G,S,k)$ can be extended to a solution of $\mathcal{I}_{i+1}$ by taking its union with $T_Z$.
Thus if $\mathcal{I}_{i+1}$ is a NO-instance, the algorithm cannot find a solution. Otherwise, let $T$ be a solution
to $\mathcal{I}_{i+1}$ with maximum possible intersection with $Z_{i+1}$.
We claim that the algorithm finds a solution in the branch $T_Z = T \cap Z_{i+1}$.
Indeed, then $(G,S,k,Z)$ is a maximal YES-instance to \dsfes{} and the algorithm $\mathcal{R}$ cannot return IGNORE.
Thus we obtain a \sfesshort{} YES-instance $(G',S',k')$,
and the algorithm from Theorem \ref{thm:small-alg} finds a solution.
\endproof

The proof of Theorem \ref{thm:reduce}
consists of a set of polynomial-time {\em{proper reductions}} (in the sense of Definition \ref{def:reduce}), each either decreasing $|V(G)|$ or decreasing $|E(G)|$ while not changing $|V(G)|$.
Some reductions may result with an IGNORE answer,
in which case the answer is immediately returned from this branch.
Note that in this case the last property of Definition \ref{def:reduce} implies
that all \dsfes{} instances in the current sequence of reductions are not maximal YES-instances.
We assume that at each step, the lowest--numbered applicable reduction is used.
If no reduction is applicable, we claim that $|S| = O(k|Z|^2)$.

We start with an obvious reduction. Note that if it is not applicable,
every edge in $S$ is contained in some simple cycle.
\begin{reduction}\label{red:bridges}
  Remove all bridges and all connected components not containing any edge from $S$.
\end{reduction}

\subsection{The outer--abundant lemma}\label{ssec:outer}

In this section we consider an instance of \dsfes{} $(G,S,k,Z)$,
where $G=(V,E)$.
We assume that Reduction \ref{red:bridges} is not applicable,
i.e., every edge in $S$ belongs to some simple cycle.
The approach here is based on ideas from the quadratic kernel for the classical
{\sc{Feedback Vertex Set}} problem \cite{fvs:quadratic-kernel},
however, a few aspects need to be adjusted to better fit our needs.
\begin{definition}
A set $F\subseteq  V$ is called {\em outer--abundant} iff:
\begin{itemize}
\item[(a)] $G[F]$ is connected,
\item[(b)] there are no edges from $S$ in $G[F]$,
\item[(c)] there at least $10k$ edges from $S$ incident with $F$.
\end{itemize}
\end{definition}

\begin{lemma}[The outer--abundant lemma]\label{lem:88k}
Let $F$ be an outer--abundant set. If Reduction \ref{red:bridges} is not applicable, then in polynomial time one can either:
\begin{itemize}
\item find a nonempty set $X\subseteq V\setminus F$ such that the following condition is satisfied:
if there exists a solution $A$ for \sfesshort{} on $(G,S,k)$ such that $A\cap F=\emptyset$, then there exists a solution $A'$ such that $A'\cap F=\emptyset$ and $X\subseteq A'$;
\item or correctly state that any solution for \sfesshort{} on $(G,S,k)$ is not disjoint with $F$.
\end{itemize}
\end{lemma}

Before we start proving Lemma \ref{lem:88k}, let us recall a few tools used in the quadratic kernel for {\sc{Feedback Vertex Set}} \cite{fvs:quadratic-kernel}.
First, we recall the result of Gallai on finding disjoint $A$--paths.

\begin{theorem}[Gallai \cite{gallai:flower}]\label{thm:gallai}
  Let $A$ be a subset of vertices of a graph $G$. A path is called an $A$--path if its endpoints are different vertices in $A$.
  If the maximum number of vertex disjoint $A$--paths is strictly less than
  $k+1$, there exists a set of vertices $B' \subseteq V$ of size at most $2k$ intersecting every $A$--path.
\end{theorem}

Moreover, it follows from Schrijver's proof of the Gallai's theorem \cite{schrijver:maders-s-paths} that Theorem \ref{thm:gallai} can be algorithmized:
in polynomial time we can find either $(k+1)$ disjoint $A$--paths or the set $B'$.

The other theorem we need is the $2$--Expansion Lemma:

\begin{theorem}[$2$--Expansion Lemma, Theorem 2.3 in \cite{fvs:quadratic-kernel}]\label{thm:2-exp}
  Let $H$ be a nonempty bipartite graph on bipartition $(X,Y)$ with $|Y| \geq 2|X|$ and such
  that every vertex of $Y$ has at least one neighbour in $X$. Then there exists
  nonempty subsets $X' \subseteq X$, $Y' \subseteq Y$ such that $N(Y')\cap X = X'$ and one can
  assign to each $x \in X'$ two {\em{private}} neighbours $y_1^x, y_2^x \in Y'$
  (i.e., each $y \in Y'$ is assigned to at most one $x \in X'$).
  In addition, such pair of subsets $X'$, $Y'$ can be computed in polynomial
  time in the size of $H$.
\end{theorem}

{\em{Proof of Lemma \ref{lem:88k}}}.
An $S$--cycle is called {\em important} if it contains an edge from $S\cap E(F,V\setminus F)$.
A set of important cycles $\{C_1, C_2, \ldots, C_t\}$ is called a {\em $t$--flower}, if the sets of vertices $C_i \setminus F$ are pairwise disjoint.

Note that if there exists a vertex $v\in V\setminus F$ such that $|E(\{v\}, F)| \geq 2$
and $E(\{v\}, F) \cap S \neq \emptyset$, then one can take $X=\{v\}$.
Indeed, any solution disjoint with $F$ has to include $v$,
since by connectivity of $G[F]$ there is an important cycle contained in $G[F\cup \{v\}]$ passing through $v$ via at least one edge from $S$.
Thus we can assume that each vertex from $V\setminus F$ is connected to $F$ by a number of edges (possibly zero) not belonging to $S$ or by a single edge from $S$.

Now we prove that in polynomial time we can find one of the following structures:
either a $(k+1)$--flower, or
a set $B$ of at most $3k$ vertices belonging to $V\setminus F$ such that each important cycle passes through at least one of them (further called a {\em $3k$--blocker}).

Let $\mathcal{C}$ be the set of those important cycles, which contain exactly two edges between $F$ and $V\setminus F$ (intuitively, visiting $F$ only once).
Note that due to connectedness of $G[F]$, a set is a $3k$--blocker iff any cycle from $\mathcal{C}$
passes through at least one of its elements.
For $C\in \mathcal{C}$ let us examine these two edges between $F$ and $V\setminus F$.
As $C$ is important, one or two of them belong to $S$. We say that such a cycle is of type I iff exactly one of these edges belong to $S$ and of type II otherwise.

Firstly, we sort out the type I cycles. We remove $F$ from the graph and replace it with two vertices $s$ and $t$. Also we add edges incident with $s$ and $t$ --- for every $vw \in E(F, V\setminus F)$
with $v \in F$, we add edge $sw$ if $vw \in S$ and edge $wt$ otherwise.
By a simple application of the vertex max--flow algorithm and Menger's theorem
one obtains either a vertex--disjoint set of paths between $s$ and $t$ of cardinality $k+1$, or a set of at most $k$ vertices such that each such a path passes through at least one of them.
Returning to the original graph transforms each path between $s$ and $t$ into a type I cycle, so we have found either a $(k+1)$-flower or a $k$-blocker of type I cycles.

Now, we deal with the type II cycles.
Let $J \subseteq V \setminus F$ be the set of vertices that are connected to $F$ by an edge from $S$.
We remove temporarily $F$ from the graph and apply Theorem \ref{thm:gallai} to the set $J$.
Note that a set of $k+1$ vertex--disjoint $J$--paths correspond to a $(k+1)$--flower, and the set $B'$ is a $2k$--blocker of type II cycles.

Using both of these methods we obtain either a $(k+1)$--flower or, by taking a union of blockers, a $3k$--blocker. Note that both algorithms run in polynomial time.

The existence of a $(k+1)$--flower immediately shows that a solution $A$ disjoint with $F$ does not exists, as each cycle belonging to a flower has to include at least one vertex from $A$.
Thus we are left only with the case of a $3k$-blocker. Let us denote it by $B$.

Let us examine $G[V\setminus(F\cup B)]$. Let $H=(V_H,E_H)$ be any of its connected components. Note that $H$ is connected to $F$ by a number of edges (possibly zero) not belonging to $S$ or by a single edge from $S$. Indeed, otherwise, due to connectedness of $H$ and of $G[F]$, there would be an important cycle contained in $G[F\cup V_H]$ not blocked by $B$. Using observation from the second paragraph of this proof, we may assume that each vertex from $B$ is connected to $F$ by at most one edge from $S$. So there are at most $3k$ edges from $S$ between $B$ and $F$. As there are at least $10k$ edges from $S$ between $F$ and $V\setminus F$, we have at least $7k$ connected components of $G[V\setminus(F\cup B)]$ connected to $F$ by a single edge from $S$.

We call a component $H$ {\em easy} if there is an $S$--cycle fully contained in $H$. If the number of easy components is larger than $k$, there is more than $k$ vertex--disjoint $S$--cycles, so $A$ does not exist. Thus we may assume that there are at least $6k$ non--easy components connected to $F$ by a single edge from $S$. We call them {\em tough} components.

Let $H=(V_H,E_H)$ be a tough component. Observe that $N(V_H)\cap B\neq \emptyset$. Indeed, otherwise the only edge between $V_H$ and $V\setminus V_H$ would be the edge from $S$ connecting $H$ with $F$ and thus a bridge sorted out by Reduction \ref{red:bridges}.

Let $T$ be the set of tough components. We construct a bipartite graph $(B\cup T,E_{exp})$ such that $vH\in E_{exp}$ iff $v\in N(V_H)$.
Note that due to observation in the previous paragraph, $(B\cup T,E_{exp})$ satisfy assumptions of Theorem \ref{thm:2-exp},
as $|T|\geq 6k = 2\cdot 3k \geq 2|B|$. So we have nonempty sets $X\subseteq B$ and $Y\subseteq T$ such that 
for every $v\in B$ there are two {\em{private}} tough components $H_{v,1},H_{v,2}\in Y$ with $v\in N(V_{H_{v,i}})$ for $i=1,2$ and $B\cap\bigcup_{H\in Y} N(H)=X$.

Let $v\in X$. Note that due to the connectedness of $H_{v,i}$ and $G[F]$, there is an $S$--cycle $C_v$ passing through $v$ --- it goes from $v$ to $H_{v,1}$, then to $F$ through an edge from $S$, then to $H_{v,2}$ through an edge from $S$ and back to $v$.
Assume that $A$ is a solution to the \sfesshort{} on $(G,S,k)$ and $A\cap F=\emptyset$.
We see that cycles $C_v$ for $v\in X$ form a $|X|$--flower, so there are at least $|X|$ vertices in
$A\cap (X\cup \bigcup_{H\in Y}V_H)$. On the other hand, each $S$--cycle passing through any vertex from $X\cup \bigcup_{H\in Y}V_H$ passes through a vertex from $X$. Indeed, each $H\in Y$ is connected to $V\setminus V_H$ with a single edge incident with $F$ and a number of edges incident with $X$. Hence each cycle passing through a vertex from $\bigcup_{H\in Y}V_H$ is fully contained in some $H\in Y$ or goes from some $H\in Y$ to $X$. As each $H\in Y$ is non--easy, cycles not incident with $X$ do not contain any edge from $S$.

These observations prove that if we construct
$$A' = \left(A \setminus \bigcup_{H\in Y}V_H\right) \cup X,$$
$A'$ will be still a solution to \sfesshort{}. Thus the set $X$ satisfies
all the required conditions.\endproof

Lemma \ref{lem:88k} allows us to greedily assume $X$ is in the solution we are looking for
(after we ensure that it is disjoint with $F$),
and either take it into the solution
(if $X\cap Z = \emptyset$) or return IGNORE (if $X\cap Z\neq \emptyset$).
As a direct application of Lemma \ref{lem:88k}, we obtain the following reduction rule. Note that if it is not applicable,
there are at most $10k|Z|$ edges from $S$ incident with $Z$.

\begin{reduction}\label{red:big-deg}
  Let $v \in Z$ be a vertex that is incident to at least $10k$ edges from $S$.
  Apply Lemma \ref{lem:88k} to the outer--abundant set $F=\{v\}$,
  If a set $X$ is returned and $X \cap Z = \emptyset$, we remove $X$ and decrease $k$ by $|X|$, otherwise
  we return IGNORE.
\end{reduction}

\subsection{Bubbles}

Recall that our goal is to reduce the size of $S$.
After Reduction \ref{red:big-deg}, there are $O(k|Z|)$ edges from $S$ incident with $Z$.
Thus, we need to care only about $S \cap E(G[V\setminus Z])$.

As $Z$ is a feasible solution to \sfesshort{} on $(G,S,|Z|)$,
every edge from $S\cap E(G[V \setminus Z])$ has to be a bridge in $G[V\setminus Z]$.
After removing those bridges $G[V\setminus Z]$ becomes an union of connected components not having
any edge from $S$. We call each such a component a {\em bubble}.
Denote the set of bubbles by $\Dd$. On $\Dd$ we have a natural structure of a graph $H=(\Dd, E_\Dd)$,
where $IJ\in E_\Dd$ iff components $I$ and $J$ are connected by an edge from $S$.
As $Z$ is a solution, $H$ is a forest and each
$I,J$ connected in $H$ are connected in $G$ by a single edge from $S$.

Consider $I\in \Dd$. Denote the set of vertices of $I$ by $V_I$.
Note that if at most a single edge leaves $I$ (that is, $|E(V_I, V \setminus V_I)| \leq 1$) then $V_I$ would be removed while processing Reduction \ref{red:bridges}.
The following reduction sorts out bubbles with exactly two outgoing edges and later
we assume that for every $I\in \Dd$ we have $|E(V_I,V\setminus V_I)|\geq 3$.
\begin{reduction}\label{red:deg2}
Let us assume that $|E(V_I,V \setminus V_I)|=2$ and let $\{u,v\}=N(V_I)$ (possibly $u=v$).
Each cycle passing through a vertex from $V_I$ is either fully contained in $I$ and
thus non--$S$--cycle, or exits $V_I$ through $u$ and $v$.
We remove $V_I$ from the graph and replace it with a single edge $uv$, belonging to $S$ iff any one of the two edges in $E(V_I, V\setminus V_I)$ is in $S$.

If the addition of the edge $uv$ lead to a multiple edge or a loop, we immediately resolve it:
\begin{itemize}
\item If $uv$ is a loop and $uv \notin S$, we delete it.
\item If $uv$ is a loop and $uv \in S$, we return IGNORE, as the fact that $Z$ is a solution to $(G,S,|Z|)$ implies that $u \in Z$.
\item If $uv$ is a multiple edge and no edge between $u$ and $v$ is from $S$, we delete the new edge $uv$.
\item If $uv$ is a multiple edge and one of the edges between $u$ and $v$ is $S$, we first note that, since $Z$ is a solution to $(G,S,|Z|)$, $u$ or $v$ is in $Z$. If both are in $Z$, we return IGNORE, otherwise we delete $\{u,v\} \setminus Z$ from the graph and decrease $k$ by one.
\end{itemize}
\end{reduction}

\input{fig2}

We are now left with bubbles that have at least three outgoing edges. We classify those bubbles according to the number of edges that connect them to other bubbles, that is $\deg_H(I)$.
\begin{definition}
We say that a bubble $I\in \Dd$ is
\begin{itemize}
\item[(a)] {\em a solitary bubble} if $\deg_H(I)=0$,
\item[(b)] {\em a leaf bubble} if $\deg_H(I)=1$,
\item[(c)] {\em an edge bubble} if $\deg_H(I)=2$,
\item[(d)] {\em an inner bubble} if $\deg_H(I)\geq 3$.
\end{itemize}
Denote by $\Dd_s, \Dd_l, \Dd_e, \Dd_i$ the sets of appropriate types of bubbles.
\end{definition}

We show that we can do some reductions to make following inequalities hold:
$$ |\Dd_l|=O(k|Z|^2), \quad\quad |\Dd_i|<|\Dd_l|, \quad\quad |\Dd_e|<3(|Z|+k)+|\Dd_i|+|\Dd_l|.$$
Note that these conditions imply that $|\Dd \setminus \Dd_s|=O(k|Z|^2)$.
As edges of $H$ create a forest over $\Dd \setminus \Dd_s$, this bounds the number of edges from $S$ not incident with $Z$
by $O(k|Z|^2)$, as desired.

\begin{lemma} \label{lem:ddi} $|\Dd_i| < |\Dd_l|$. \end{lemma}
  \begin{proof}
    As $H$ is a forest, then $|E_\Dd|<|\Dd|-|\Dd_s|=|\Dd_i|+|\Dd_e|+|\Dd_l|$. Moreover, $2|E_\Dd|=\sum_{I\in \Dd} \deg_H(I) \geq 3|\Dd_i|+2|\Dd_e|+|\Dd_l|$. Therefore $|\Dd_l|>|\Dd_i|$.
  \end{proof}

\begin{reduction} \label{red:edge_bub}If $|\Dd_e| \geq 3(|Z| + k) + |\Dd_i| + |\Dd_l|$, then return IGNORE. \end{reduction}

\begin{lemma} Reduction \ref{red:edge_bub} is a proper reduction. \end{lemma}
\begin{proof}
We first show that the number of edge bubbles not adjacent to any other edge bubble in $H$ is at most $|\Dd_i|+|\Dd_l|$.
Let us root each connected component of $H$ in an arbitrary leaf and for any bubble $I\in \Dd_e$ let
$\varphi(I)$ be the only child of $I$. Observe that this mapping is injective and maps the set of edge bubbles isolated in $H[\Dd_e]$ into $\Dd_i \cup \Dd_l$.

We now prove by contradiction if the reduction is applicable, every feasible solution of $(G,S,k)$ contains a vertex from $Z$.
As edge bubbles have degree $2$ in $H$, $H[\Dd_e]$ is a set of paths of non-zero length
and isolated vertices.
There are at least $3(|Z|+k)$ vertices contained in the paths.
Let $M$ be a maximal matching in $H[\Dd_e]$. If a path contains $l$ vertices (for $l\geq 2$), it has a matching of cardinality $\lfloor\frac{l}{2}\rfloor\geq \frac{l}{3}$.
Therefore, in $H[\Dd_e]$ we have a matching of cardinality at least $|Z|+k$. Let us examine an arbitrary $IJ\in M$. Recall that at least three edges leave each bubble.
As each of $V_I$, $V_J$ is adjacent to two other bubbles by single edges, it has to be connected to $Z$ as well.
Choose $u_I,u_J$ --- vertices from $Z$ such that $u_I\in N(V_I)$ and $u_J\in N(V_J)$.
We see that there is a path from $u_I$ to $u_J$ passing through an edge from $S$: it goes from $u_I$ to $I$, then to $J$ through an edge from $S$, and then to $u_J$.
As $M$ is a matching, such paths are vertex--disjoint for all $IJ\in M$, except for endpoints $u_I$ and $u_J$.

If we have a solution disjoint with $Z$ of cardinality at most $k$, there are at least $|Z|$ pairs $IJ\in M$, where neither $I$ nor $J$ contains a vertex from the solution.
Now we construct a graph $P=(Z,E_P)$ such that $u_Iu_J\in E_P$ if $u_I,u_J$ have been chosen for some $IJ\in M$, where $I$ and $J$ are solution--free.
We prove that $P$ has to be a forest.
Indeed, otherwise there would be a cycle in $P$ --- and by replacing each edge from it by associated path, we construct an $S$-cycle in $G$ (as paths in which edges from $E_P$ originated are vertex--disjoint).
This cycle does not contain any vertex from the solution, as it passes only through $Z$ and solution--free bubbles.
However, as $|E_P|\geq |Z|$, $P$ cannot be a forest; the contradiction ends the proof.
\end{proof}

\subsection{The leaf bubble reduction}\label{ssec:leaf}
We are left with the leaf bubbles and we need to show reductions that lead to $|\Dd_l| = O(k|Z|^2)$. We do this by a single large reduction described in this subsection.
It proceeds in a number of steps. Each step either returns IGNORE (thus ending the reduction) or --- after, possibly, modifying $G$ --- passes to the next step.
Each step is not a standalone reduction, as it may increase $|E(G)|$. However, if the reduction below is fully applied, it either returns IGNORE or reduces $|V(G)|$.

Let $I$ be a leaf bubble. As there are at least three edges leaving $I$,
each leaf bubble is connected to $Z$ by at least two edges.
We begin with a bit of preprocessing:
\begin{step}\label{step:addedges}
As long as there are two vertices $v,v'$ in $Z$ with $vv' \notin E$, and at least $k+1$ bubbles, each connected to both $v$ and $v'$ by edges not in $S$,
we add an edge $vv'$ to $E$, with $vv' \notin S$.
\end{step}

\begin{lemma} The output $(G',S,k)$ of Step \ref{step:addedges} and the input $(G,S,k)$, as \sfesshort{} instances, have equal sets of feasible solutions. \end{lemma}

\begin{proof}
Obviously any solution  to $(G',S,k)$ is a solution to $(G,S,k)$, as we only added edges (and thus only added potential $S$--cycles).
On the other hand, suppose we have a solution $T$ to $(G,S,k)$ which is not a solution to $(G',S,k)$.
Then there is some $S$--cycle $C$ in $(G',S,k)$ not passing through any vertex of $T$.
$C$ has to pass through the edge $vv'$ (otherwise it would also be an $S$--cycle in $(G,S,k)$).

As $|T| \leq k$, there is at least one bubble $I_j$ which is disjoint with $T$.
Thus we can find a simple path $P$ connecting $v$ and $v'$, the interior vertices of which are all in $I_j$.
Note that as $I_j$ is a bubble and the edges to $v$ and $v'$ were not in $S$, $P$ does not contain any edge from $S$.
Consider the cycle $C'$ (not necessarily simple) in $G$ which is formed by replacing the edge $vv'$ in $C$ by the path $P'$.
As $C$ was an $S$--cycle, there is some edge $e \in S \cap C$.
This edge is visited by $C'$ exactly once --- as we took out $vv' \notin S$ and added edges from $P$, which is disjoint with $S$.
If we consider the graph spanned by edges from $C'$, the edge $e$ is not a bridge in this graph, as the endpoints of $e$ are connected
by $C' \setminus \{e\}$. Therefore, $e$ lies on some simple cycle contained in $C'$, a contradiction.
\end{proof}

Now for each bubble $I$ with vertex set $V_I$ we choose arbitrarily two of the edges connecting it to $Z$: $e_I$ and $e_I'$. Additionally assume that
if $S \cap E(V_I, Z) \neq \emptyset$ then $e_I \in S$.
Let $v_I$ and $v_I'$ be the endpoints of $e_I$ and $e_I'$ in $Z$ (possibly $v_I = v_I'$). We say that a bubble $I$ is {\em associated} with vertices $v_I, v_I'$.
If two leaf bubbles $I_1,I_2$ are connected in $H$ (form a $K_2$ in $H$), by an edge $e_{I_1I_2}\in S$, we call them a {\em bubble--bar}.
The proofs of the following lemmata proceed along lines similar to the proof of correctness for Reduction \ref{red:edge_bub}:

\begin{lemma} \label{lemma:edgesIZS}
  If there are at least $|Z|^2(k+2)$ leaf bubbles $I$
  such that $e_I \in S$ then every feasible solution of \sfesshort{} on $(G,S,k)$
  contains a vertex from $Z$.
\end{lemma}

\begin{proof}
  By the Pigeonhole Principle, there exist $v,v' \in Z$ associated with at least $k+2$ of the considered leaf bubbles.
  If $v=v'$, there are $k+2$ $S$--cycles sharing only $v$, each constructed from a different bubble $I$ by closing a path contained in $I$ with edges $e_I$ and $e_I'$. Therefore, $v$ needs to be part of any feasible solution.
  If $v \neq v'$, one can similarly choose a path between $v$ and $v'$ which contains an edge from $S$ through each bubble $I$. These $k+2$ paths are vertex--disjoint apart from $v$ and $v'$, so any feasible solution disjoint with $Z$ leaves at least two of them solution--free. These two paths can be arranged into a solution--free $S$--cycle, so any feasible solution is not disjoint with $Z$, as it contains $v$ or $v'$.
\end{proof}

\begin{lemma}\label{lemma:edgesZZS}
  If there are at least $|Z|^2(k+1)$ leaf bubbles $I$
  such that $v_Iv_I' \in S$, then every feasible solution of \sfesshort{} on $(G,S,k)$
  contains a vertex from $Z$.
\end{lemma}

\begin{proof}
  As before, there exist $v,v' \in Z$ associated with at least
  $k+1$ considered leaf bubbles.
  These bubbles generate at least $k+1$ $S$--cycles, which are vertex--disjoint apart from $v,v'$.
  Therefore, any feasible solution needs to include $v$ or $v'$.
\end{proof}

\begin{lemma}\label{lemma:bubblebars}
  If there are at least $|Z|^2(k+2)$ bubble--bars, then
  any feasible solution of \sfesshort{} on $(G,S,k)$ contains a vertex from $Z$.
\end{lemma}

\begin{proof}
  By the Pigeonhole Principle, there exist $v, v' \in Z$ such that
  there exist at least $k+2$ bubble--bars $(I_1,I_2)$ with
  $v_{I_1} = v$ and $v_{I_2}=v'$. If $v=v'$,
  there are $k+2$ $S$--cycles having only $v$ in common (one through each bubble--bar), so any feasible solution has to contain $v$.
  If $v \neq v'$, there are $k+2$ paths connecting $v$ and $v'$ and sharing only $v$ and $v'$. Any solution disjoint with $Z$ would leave at least two of them solution--free. Then these two paths could be arranged into a solution--free $S$--cycle.
\end{proof}

The above lemmata justify our next step.

\begin{step} \label{step:first}If any of the situations from Lemmata \ref{lemma:edgesIZS}, \ref{lemma:edgesZZS} and \ref{lemma:bubblebars} occur, return IGNORE.\end{step}

Summing all the obtained bounds, we can count almost all the leaf bubbles (possibly more than once) and bound their number by $O(k|Z|^2)$. The ones that are left satisfy the following definition:
\begin{definition} A leaf bubble $I$ satisfying the following three conditions is called a {\em clique bubble}:
\begin{itemize}
\item[(a)] $G[N(V_I)\cap Z]$ is a clique not containing any edge from $S$,
\item[(b)] $I$ is connected to $Z$ by edges not belonging to $S$,
\item[(c)] $I$ is connected to a non--leaf bubble.
\end{itemize}
\end{definition}

Denote the only edge from $S$ connecting a given clique bubble $I$ with its neighbour bubble by $w_Iw_I'$, with $w_I\in V_I$.

\begin{lemma} \label{lem:clique-disjoint}If there exists a feasible solution $T$ for \sfesshort{} on $(G,S,k)$, then there exists a feasible solution $T'$, which is disjoint from all clique bubbles in $G$. Moreover, if $T$ is disjoint with $Z$, so is $T'$.\end{lemma}

\begin{proof}
Let $I$ be a clique bubble. Assume we have a feasible solution $T$, with $T \cap V_I \neq \emptyset$.
We show $T' = (T \setminus V_I) \cup \{w_I'\}$ is also a feasible solution.
Consider any $S$--cycle $C$ in $G[V\setminus T']$. This cycle has to pass through $V_I$ (possibly multiple times), or it would be an $S$--cycle in $G[V \setminus T]$, contrary to the assumption $T$ was a feasible solution.
Note that $C$ has to enter and exit $V_I$ through $N(V_I) \cap Z$, as the only vertex in $N(V_I) \setminus Z$ is $w_I'$, which is removed by $T'$.
But then $C$ can be shortened to $C'$ by replacing every part contained in $V_I$ by a single edge in $Z$ (as $N(V_I) \cap Z$ is a clique).
Now $C'$ is disjoint from $V_I$ and is an $S$--cycle due to the definition of the clique bubble. So $C'$ is an $S$--cycle in $G[V\setminus T]$, a contradiction.

Note that the only vertex we added to $T$ was $w_I'$, which does not belong to a clique bubble (it does not belong even to a leaf bubble, from property (c) in the definition of clique bubbles).
Thus we can apply this procedure inductively, at each step reducing the number of vertices in $T$ contained in clique bubbles, until none are left.
\end{proof}

Assume there is a vertex $v\in Z$ such that $v\in N(V_{I_j})$ for some distinct clique--bubbles $I_1,I_2,\ldots,I_{10k}$. We show that the set $F=\{v\}\cup \bigcup_{j=1}^{10k}V_{I_j}$
is outer--abundant in $G$.
Indeed, it is connected and due to the definition of bubbles and properties of the clique bubble definition, the subgraph $G[F]$ does not contain edges from $S$.
Moreover, there are at least $10k$ edges from $S$ incident with $G[F]$ --- these are the edges connecting bubbles $I_j$ with other bubbles, not contained in $F$ as they are non--leaf ones due to property (c).
This enables us to formulate the key step:

\begin{step}\label{step:final}
If there is a vertex $v \in Z$ which is adjacent to at least $10k$ clique bubbles, we apply Lemma \ref{lem:88k} to the set $F = \{v\} \cup \bigcup_{j=1}^{10k} V_{I_j}$.
If a set $X$ is returned and $X \cap Z = \emptyset$, we remove $X$ from the graph and decrease $k$ by $|X|$, otherwise we return IGNORE. \end{step}

Suppose there is a feasible solution $T$ to $(G,S,k)$.
Due to Lemma \ref{lem:clique-disjoint} we may assume $T$ to be disjoint with $F \setminus \{v\}$.
Thus either $T$ contains $v$, or it is disjoint with $F$, and by Lemma \ref{lem:88k} there exists a solution containing $X$.
This justifies the correctness of Step \ref{step:final}.

Now we summarize the steps made in this section
to show clearly that the number of leaf bubbles is bounded by $O(k|Z|^2)$.

Assume no reduction is applicable. Note that in the last run, the last reduction may add some edges in Step \ref{step:addedges}. Let $G'$ denote the modified graph.
Let us check that the graph $G'$ indeed has $O(k|Z|^2)$ edges from $S$:
\begin{enumerate}
\item The decomposition of $V(G) \setminus Z$ into bubbles is the same as the decomposition of $V(G') \setminus Z$ and bubbles that were inner or edge bubbles in $G$ are, respectively, inner or edge bubbles in $G'$;
\item If Step \ref{step:first} is not applicable, there are at most $|Z|^2 (k+2) - 1$ leaf bubbles connected to $Z$ by an edge from $S$, at most $|Z|^2 (k+1) - 1$ leaf bubbles associated with a pair of vertices connected with an edge from $S$, and at most $2|Z|^2 (k+2)$ leaf bubbles connected to other leaf bubbles.
\item If Step \ref{step:addedges} is not applicable, for any pair $v,v'$ of vertices in $Z$ with $vv' \notin E$ there are at most $k$ leaf bubbles adjacent to both vertices of that pair through edges not in $S$.
\item If a leaf bubble is not a clique bubble, it either is connected to a leaf bubble (forming a bubble--bar), is connected to $Z$ by an edge in $S$, has an edge from $S$ between some two of its neighbours in $Z$, or has some two neighbours in $Z$ not connected by an edge. The number of such bubbles in all four cases was estimated above. Thus, in total, there are at most $O(k|Z|^2)$ bubbles which are not clique bubbles.
\item Finally, if Step \ref{step:final} is not applicable, there are at most $(10 k - 1) |Z|$ clique bubbles.
\item Thus $|\Dd_l| = O(k|Z|^2)$, moreover $|\Dd_i| \leq |\Dd_l|$ by Lemma \ref{lem:ddi} and $|\Dd_e| \leq 3(|Z| + k) + |\Dd_i| + |\Dd_l|$ by Reduction \ref{red:edge_bub} --- thus the number of edges in $S$ not incident with $Z$ is bounded by $O(k|Z|^2)$. We added no new edges to $S$, and the number of edges in $S$ incident to $Z$ was bounded by $O(k|Z|)$ in the input graph, thus in the output graph there are $O(k|Z|^2)$ edges from $S$, as desired. 
\end{enumerate}

Thus we managed to reach the state when the number of leaf bubbles is bounded by $O(k|Z|^2)$.
As we modified only the subgraph $G[Z]$, the sets $\Dd_i$, $\Dd_e$, $\Dd_l$ remain the same
after modifications and we obtain a graph with $|S| = O(k|Z|^2)$.
This completes the description of the
$2^{O(k \log k)} n^{O(1)}$ algorithm for \sfesshort{}.

%% file: fig2.tex
  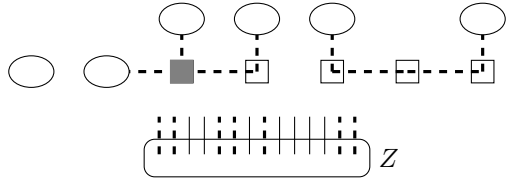
\begin{figure}[htbp]
  \begin{center}
  \begin{tikzpicture}[scale=1]
  
   \begin{scope}[yshift=-1.5cm]

   \foreach \x/\nx in {-2/-1,-1/0,1/2,2/3} 
    {
       \draw[very thick, dashed] (\x,1.4) -- (\nx,1.4);
    }
    \foreach \x in {-1,0,1,3}
    {
       \draw[very thick, dashed] (\x, 1.4) -- (\x, 2.1);
    }    
\begin{scope}[yscale=0.7]
   \foreach \x/\y in {-3/2, -2/2, -1/3, 0/3, 1/3, 3/3}
   {
      \fill[white] (\x,\y) circle (0.3);
      \draw (\x, \y) circle (0.3);
   }
   \end{scope}
   
   \foreach \x/\lx/\rx in {-1/-1.15/-0.85, 0/-0.15/0.15, 1/0.85/1.15, 2/1.85/2.15, 3/2.85/3.15}
   {
      \fill[white] (\rx, 1.25) rectangle (\rx, 1.55);
      \draw (\lx, 1.25) rectangle (\rx, 1.55);
   }   
    \fill[gray] (-1.15,1.25) rectangle (-0.85,1.55);
   
   \end{scope}

   \draw[rounded corners] (-1.5,-1.5) rectangle (1.5,-1);
   \draw (1.75,-1.25) node {$Z$};
   \foreach \x in {-1.3, -1.1, -0.5, -0.3, 0.1, 1.1, 1.3}
     \draw[very thick, dashed] (\x, -1.2) -- (\x, -0.7);
   \foreach \x in {-0.9, -0.7, -0.1, 0.3, 0.5, 0.7, 0.9}
     \draw[black] (\x, -1.2) -- (\x, -0.7); 

\end{tikzpicture}
\end{center}
\caption{Set of vertices $Z$ and a forest of bubbles. Ellipse-shaped bubbles
  represent leaf bubbles, white squares represent edge bubbles and squares
    filled with gray represent inner bubbles.  Dashed edges belong to the set $S$.}
\end{figure}

%% file: r-reduction.tex
\section{The relationship of \sfvsshort{} and terminal separation}\label{sec:red}

It is known (e.g.~\cite{sfvs:8-apx}) that in the weighted case the 
\nodemultiwayname{} problem
can be reduced to weighted \sfvsshort{} by adding a vertex $s$ 
(with infinite weight)
to the graph and connecting it to all the terminals, where $S = \{s\}$.
Here we present a modified version, adjusted to the unweighted 
parameterized setting.
Both the node and edge versions of the \multiwayname{} problem are known 
to be FPT since 2004~\cite{marx:multiway-journal}.

\begin{theorem}
\label{thm:multiway-to-sfvs}
An instance $(G,\TT,k)$ of the \nodemultiwayname{} problem can be transformed in polynomial time
into an equivalent instance $(G',S,k)$ of the \sfesname{} problem.
\end{theorem}

\begin{proof}
Let $\TT=\{v_1,\ldots,v_t\}$.
We add a set $\TT'={v_1',\ldots,v_t'}$ of $t$ vertices to the graph $G$ obtaining a new graph $G'$.
Together with the vertices from the set $\TT'$ we add a set of edges $S=\{v_iv_i': 1 \le i \le t\}$.
Moreover, we add an edge between every pair of vertices from the set $\TT'$ so that $G[\TT']$ becomes a clique.
Assume that $(G,\TT,k)$ is a YES-instance of \nodemultiwayname{} where $T \subseteq V$ is a solution.
Clearly $T$ is a solution for the instance $(G',S,k)$ of \esfvsshort{} since an $S$-cycle in $G'[(V\cup \TT') \setminus T]$
implies a path between terminals in $G[V\setminus T]$ (see Fig.~\ref{fig:redukcja}).

In the other direction, assume that $(G',S,k)$ is a YES-instance of \esfvsshort{} where $T \subseteq V$
is a set of removed vertices.
Let $T' = T \setminus \TT' \cup \{v_i: v_i' \in T\}$.
We now prove that $T'$ is a solution for the \nodemultiwayname{} instance $(G,\TT,k)$.
Clearly $|T'| \leq |T| \leq k$. Assume that there exists a path $P$ in $G[V \setminus T']$ between
terminals $v_{i_1}$ and $v_{i_2}$. In particular, this means that $v_{i_1}, v_{i_2} \notin T'$, so
$v_{i_1}, v_{i_2}, v_{i_1}', v_{i_2}' \notin T$. Thus the path $P$ together with the path $v_{i_1}v_{i_1}'v_{i_2}'v_{i_2}$
forms an $S$-cycle that is not hit by $T$, a contradiction.
\end{proof}

\input{fig1}

%% file: fig1.tex
  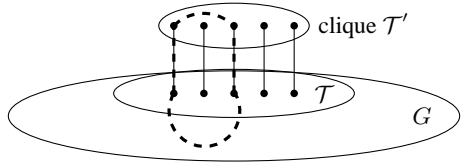
\begin{figure}[htbp]
  \begin{center}
  \begin{tikzpicture}[scale=1,font=\small]
  \foreach \x in {-0.8, -0.4, 0, 0.4, 0.8}
  {
    \draw (\x, 0.6) -- (\x, 1.5);
    \fill[black] (\x, 1.5) circle (0.05);   
  }

   \begin{scope}[yscale=0.3]
      \draw (0,5) circle (1.0);
   \end{scope}
   \draw[right] (1.0, 1.5) node {clique $\TT'$};

   \draw[dashed,very thick] (-0.8,0.6) arc (-210:45:0.47);
   \draw[very thick, dashed] (-0.8, 0.6) -- (-0.8, 1.5);
   \draw[very thick, dashed] (0, 0.6) -- (0, 1.5);
   \draw[dashed,very thick] (-0.8,1.5) arc (135:45:0.56);

  \draw (2.5,0.3) node {$G$};
  \draw (1.2,0.6) node {$\TT$};
  \begin{scope}[yscale=0.2]
  \draw (0,1.5) circle (3.0);
  \draw (0,3) circle (1.6);
  \end{scope}
  
  \foreach \x in {-0.8, -0.4, 0, 0.4, 0.8}
  {
    \fill[black] (\x, 0.6) circle (0.05);
  }

\end{tikzpicture}
\end{center}
\caption{Reduction used in Theorem~\ref{thm:multiway-to-sfvs}}.
\label{fig:redukcja}
\end{figure}

%% file: r-conclusions.tex
\section{Conclusions}\label{sec:conc}

In this paper we presented a fixed-parameter algorithm for \sfvsname{}, making extensive use of recently discovered tools in parameterized complexity such as iterative compression, Gallai's theorem and $c$-Expansion Lemma. To settle down the exact parameterized complexity of \sfvsshort{}, one question remains: does this problem admit a polynomial kernel?
Kratsch and Wahlstr\"{o}m \cite{oct-kernel} very recently gave a polynomial kernel for {\sc{Odd Cycle Transversal}}, using a compact representation of special classes of matroids.
These tools may be useful in our problem as well.

Second, can we improve the time complexity of our algorithm? In particular: is there a $c^k n^{O(1)}$ algorithm for \sfvsname{}? Or maybe we can show that this is unlikely,
using the recent framework of Lokshtanov et al. \cite{lms:superexp}?